# Large irreversibility field in nanoscale C-doped $MgB_2$/Fe tape conductors

Yanwei Ma [a)], Xianping Zhang

*Applied Superconductivity Lab., Institute of Electrical Engineering, Chinese Academy of Sciences,*
*P.O. 2703, Beijing 100080, China*

S. Awaji

*High Field Laboratory for Superconducting Materials, Institute for Materials Research,*
*Tohoku University, Sendai 980-8577, Japan*

Zhaoshun Gao, Dongliang Wang, Zhengguang Yu

*Applied Superconductivity Lab., Institute of Electrical Engineering, Chinese Academy of Sciences,*
*P.O. 2703, Beijing 100080, China*

G. Nishijima, K. Watanabe

*High Field Laboratory for Superconducting Materials, Institute for Materials Research,*
*Tohoku University, Sendai 980-8577, Japan*

**Abstract**

We investigated the effect of nanoscale-C doping on the critical current density $J_c$ and irreversibility field $B_{irr}$ of Fe-sheathed $MgB_2$ tapes prepared by the *in-situ* powder-in-tube method. The tapes were heat treated at 600-950°C for 1 h. Higher values of $J_c$ and $B_{irr}$ were seen for 5 at.%C-doped $MgB_2$ tapes at higher sintering temperatures, where substantial substitution of boron for carbon occurred. The C-doped samples sintered at 950°C showed the highest $B_{irr}$, for example, at 4.2 K, the $B_{irr}$ reached 22.9 T. In particular, at 20 K, $B_{irr}$ for the C-doped tape achieved 9 T, which is comparable to the upper critical field of the commercial NbTi at 4.2 K. This role of nano-sized C particles can be very beneficial in the fabrication of $MgB_2$ tapes for magnetic resonance imaging applications at 20 K.

[a)] electronic mail: ywma@mail.iee.ac.cn

**Introduction**

Compared to conventional metallic superconductors, $MgB_2$ has advantages of high transition temperature ($T_c$) and low raw material costs of both B and Mg. It is believed that $MgB_2$ can be used with a convenient cryocooler as a conductor for a cryogen-free magnet at elevated temperatures of around 20 K [1]. In particular, recent studies found a significant upper critical field $B_{c2}$ enhancement, $B_{c2}^{\perp}$ (4.2 K) ≈ 35 T and $B_{c2}^{//}$ (4.2 K) ≈ 51 T, in C-alloyed films [2]. Such critical field properties exceed those of any $N_b$-based conductor at any temperature, suggesting that $MgB_2$ could also be a replacement for the well-known $Nb_3Sn$ as a high field magnet conductor.

The method commonly used to fabricate $MgB_2$ tape is the powder-in-tube technique [3]. So far, enormous efforts have been directed towards improving $B_{c2}$ and in-field critical current density ($J_c$) through development and application of various methods for fabrication of technically usable materials, such as chemical doping [4-8], irradiation with heavy ions [9], and various processing techniques [10-11]. Among all the methods, chemical doping is the most promising way for applications. Since $MgB_2$ has a relatively large coherence length and small anisotropy, the fluxoids to be pinned are string-like and amenable to pinning by inclusions and precipitates in the grains.

Dou et al. [4] have reported that SiC doping can significantly improve the irreversibility field ($B_{irr}$) and $J_c$. Kumakura et al. [6] used hydride-based $MgB_2$ powder with SiC dopants in the *in-situ* process, enhancing $B_{irr}$ from 17 T to 22.5 T at 4.2 K. SiC-doped $MgB_2$ has also been attempted by Sumption et al.[7] in metal sheathed strands reaching $B_{c2}$ values up to 33 T. Since C element is widely recognized to enter the structure through replacing boron, it is expected that the carbon, which has one more electron than boron, will donate electrons to the σ band. Thus, C doping of $MgB_2$ is believed to be a quite useful means of alloying $MgB_2$ for enhancing $B_{c2}$ and flux pinning [12-13]. Recently, we have demonstrated that a $J_c$ enhancement by more than one order of magnitude in high magnetic fields can be easily achieved through doping the $MgB_2$/Fe tape with C nanoparticles. The highest $J_c$ value of tapes was achieved in the 5 at.% nano-C addition [14]. Therefore, we expected that transport measurements in higher fields on C-doped samples would provide additional useful information for understanding the $J_c$ and $B_{irr}$ behaviors of $MgB_2$ tapes. In this work, the effect of sintering temperature on $J_c$-B property was investigated, and high-field resistive transitions were used to demonstrate relatively large values of $B_{irr}$ of C-added $MgB_2$ tapes under various heating conditions.

**Experimental**

Powders of magnesium (99.8%, -325 mesh), amorphous boron (99.99%, 2-5μ) and carbon nanoparticle powders (20-30 nm, amorphous) were used for the fabrication of tapes by the *in-situ* powder-in-tube method. The nano-C concentration was fixed to be 5 at.%. Details of the tape fabrication procedure have been described elsewhere [14]. The sheath materials chosen for this experiment were commercially available pure Fe. The mixed powder was filled into a Fe tube of 8 mm outside diameter and 1.5 mm wall thickness in air. After packing, the tube was rotary swaged and drawn to wires of 1.5 mm in diameter. The wires were subsequently rolled to tapes of ~3.2 × 0.5 mm. Short samples (~ 4 cm each), cut from the tapes, were heated in flowing *Ar* at temperatures ranging from 600 to 950°C for 1 h. Undoped tapes were also prepared under the same conditions for use as reference samples.

In order to precisely evaluate the $B_{irr}$ values at elevated temperatures, resistive transition measurements were made on 1.5 cm long tapes at the High Field Laboratory for Superconducting Materials (HFLSM) in Sendai. The resistance versus temperature curves were measured in magnetic fields up to 17 T by a four-probe method using a 20 T superconducting magnet. Four-point transport resistive measurements in high magnetic field up to 26 T at 4.2-10 K were also made using a 28 T hybrid magnet. The distance between the voltage taps was 5 mm. The applied current was 100 mA. The temperature of the tapes was changed by the combination of He gas cooling and heating, and the temperature of the tape was monitored with a thermometer attached directly to the tape. A magnetic field was applied parallel to the tape surface and perpendicular to the current flow. Values of $B_{irr}$ were obtained taking the 10% points of the resistive transition. The transport $J_c$ at 4.2 K and its magnetic field dependence were evaluated by a standard four-probe technique with a criterion of 1 μV/cm. Magnetic fields were applied parallel to the tape surface by employing either a newly developed 18 T cryogen-free superconducting magnet [15] or the 28 T hybrid magnet at the HFLSM.

**Results and discussion**

Figure 1 shows the $J_c$ at 4.2 K in magnetic fields up to 18 T for 5% C-doped tapes that were sintered at different temperatures ranging from 600 to 950°C. The $J_c$ values of an undoped tape heated at 800°C are also included as a standard. Only data above 5 T are shown, because at lower field region, $I_c$ was too high to be measured. From Fig. 1, we immediately notice that the sintering temperature has a significant effect on the $J_c$-B performance. The $J_c$ values of doped samples increased systematically with increasing sintering temperatures. All the C-doped samples sintered at different temperatures exhibited superior $J_c$ values compared to the pure tape in measuring fields of up to 18 T. The tape sintered at 950°C revealed the highest $J_c$ values with excellent $J_c$-B performance compared to all other samples: at 4.2 K, the transport $J_c$ reached $2.11 \times 10^4$ A/cm$^2$ at 10 T, over an order of magnitude larger than for the undoped sample. Furthermore, for C-added tapes, the field dependence of the $J_c$ became weaker with increasing heat treatment temperature. This behavior is quite in contrast to the SiC-doped case [16], where the $J_c$-B curves were independent of the sintering temperature. In fact, it is recognized that the higher the sintering temperature is, the larger the proportion of C that is substituted for B in MgB$_2$ [12-13, 17]. Therefore, our C-doped samples heated at higher temperatures showed a better field performance and higher $J_c$, indicating that a higher annealing temperature promotes more C substitution reaction for B, thus enhancing flux pinning and improving the high-field $J_c$.

To identify the critical current properties in higher magnetic fields above 18 T, transport $J_c$ measurements were carried out by using the hybrid magnet at the HFLSM in Sendai. Figure 2 shows the magnetic-field dependence of the transport $J_c$ at 4.2 K for pure and 5% C-doped MgB$_2$/Fe tapes. Again, both $J_c(B)$ values of C-doped MgB$_2$ tapes were higher than for the pure MgB$_2$ sample, and the tape heated at 950°C still exhibited the largest $J_c$ value. Specifically, C-doped tapes showed much smaller dependence of $J_c$ on magnetic field due to the C addition, which can be explained by the introduction of effective pinning centers in higher fields. However, the $J_c$ of the undoped tape decreased rapidly with increasing the magnetic field. In addition, the $J_c$ value of the pure sample fell to 10 A/cm$^2$ at 16 T. For the C-doped tape sintered at 950°C, on the other hand, $J_c$ higher than 10 A/cm$^2$ was observed at 23 T. Clearly, doping with C significantly improved the $B_{irr}$ values of $MgB_2$ tapes well above 20 T.

In order to precisely evaluate the $B_{irr}$ values at elevated temperatures, we performed the resistance versus temperature measurements on the pure and C-added $MgB_2$ tapes in various magnetic fields by the four-probe resistive method. With increasing the magnetic field, both the onset and zero resistive points of the superconducting transition curve shifted towards low temperatures for $MgB_2$ tapes, as typically shown in Fig. 3. The $T_c$ of the pure sample was 36.5 K. For the C-doped tape sintered at 800°C, $T_c$ decreased by 1.8 K. By contrast, $T_c$ was depressed to about 34.2 K for the doped sample heated at 950°C. This indicates that the extent of the C substitution reaction increases with increasing sintering temperature, resulting in $T_c$ depression, which is in agreement with a recent report [17]. As we can see, although the pure sample has a higher $T_c$ value at zero field, $T_c$ was depressed more severely in the un-doped sample compared to the C-doped one by the applied magnetic field.

Figure 4 shows the temperature dependence of $B_{irr}$ for the pure and C-added tapes, obtained from the 10% values of their corresponding resistive transitions. Clearly, the $B_{irr}$ of the C-doped tapes increased more rapidly with decreasing temperature than that of the undoped one. It is noted that both the $B_{irr}$ curves for doped tapes showed a crossover with the pure sample at higher temperature, for instance, the crossover at 26 K for an 800°C curve while at 29 K for the 950°C curve. Although the C addition introduced a degradation of the $T_c$, the C-doped tapes show a higher $B_{irr}$ in the low temperature region. In fact, the C-doped samples sintered at 950°C showed the highest $B_{irr}$.

Figure 4 suggests that the higher sintering temperature led to a higher $H_{irr}$ value in C-doped $MgB_2$ tapes, a result indicating that a degree of carbon substitution improves with increasing heat treatment temperature. This increase in the C substitution in the B sites from higher sintering temperatures, results in an increase in the impurity scattering in the σ band and a consequent improvement in the $B_{c2}$ value.

It is also interesting to note from Fig.4 that at 4.2 K, $B_{irr}$ for the C-doped tape heated at 950°C reached 22.9 T, compared to 16 T for the undoped one. This is consistent with $B_{irr}$ data being determined by the transport $J_c(B)$ measurements (see Fig.2). The $B_{irr}$ value of 22.9 T, which is even slightly higher than that of SiC-added $MgB_2$ tapes using $MgH_2$+B powders [6], was comparable to the $B_{c2}$ of a conventional bronzed-processed

Nb$_3$Sn conductor. Most importantly, at 20 K, the B$_{irr}$ achieved 9 T for C-doped tapes heated at 950°C, which was comparable to the B$_{c2}$ at 4.2 K of commercial NbTi conductors [18]. This result clearly demonstrates that magnetic resonance imaging magnet made by NbTi wires and operated at 4.2 K, can be replaced with a convenient cryogen-free magnet operated at around 20 K fabricated with MgB$_2$ conductors.

**Conclusions**

In summary, we have synthesized nano-C doped MgB$_2$/Fe tapes by applying heat treatments at 600-950°C. Higher values of J$_c$(B) and B$_{irr}$ were seen for C-doped MgB$_2$ tapes at higher sintering temperatures, where substantial substitution of boron for carbon occurred. In particular, the C-added tapes showed the B$_{irr}$ value comparable to that of commercial NbTi at 20 K. This role of nano-sized C particles can be very beneficial in the fabrication of MgB$_2$ tapes for magnetic resonance imaging applications at 20 K.

**Acknowledgments**

We thank Yulei Jiao, Ling Xiao, Haihu Wen, Liye Xiao and Liangzhen Lin for their help and useful discussions. This work is partially supported by the National Science Foundation of China (NSFC) under Grant Nos.50472063 and 50572104 and National "973" Program (Grant No. 2006CB601004).

**Captions**

Figure 1 Transport $J_c$ at 4.2 K in magnetic fields up to 18 T for 5 at.% C-doped $MgB_2$ tapes sintered at various temperatures. The $J_c$ values of an undoped tape heated at 800°C are also included as a standard.

Figure 2 Transport $J_c$-B curves at 4.2 K in the high magnetic fields of pure and 5 at.% C-doped $MgB_2$ tapes.

Figure 3 The typical resistance vs temperature curves for the pure and 5 at.% C-doped $MgB_2$ tapes heated at 800°C.

Figure 4 $B_{irr}$ values as a function of the temperature for the pure and 5 at.% C-doped $MgB_2$ tapes. The $B_{irr}$ values were defined as the 10% points of the resistive transition.

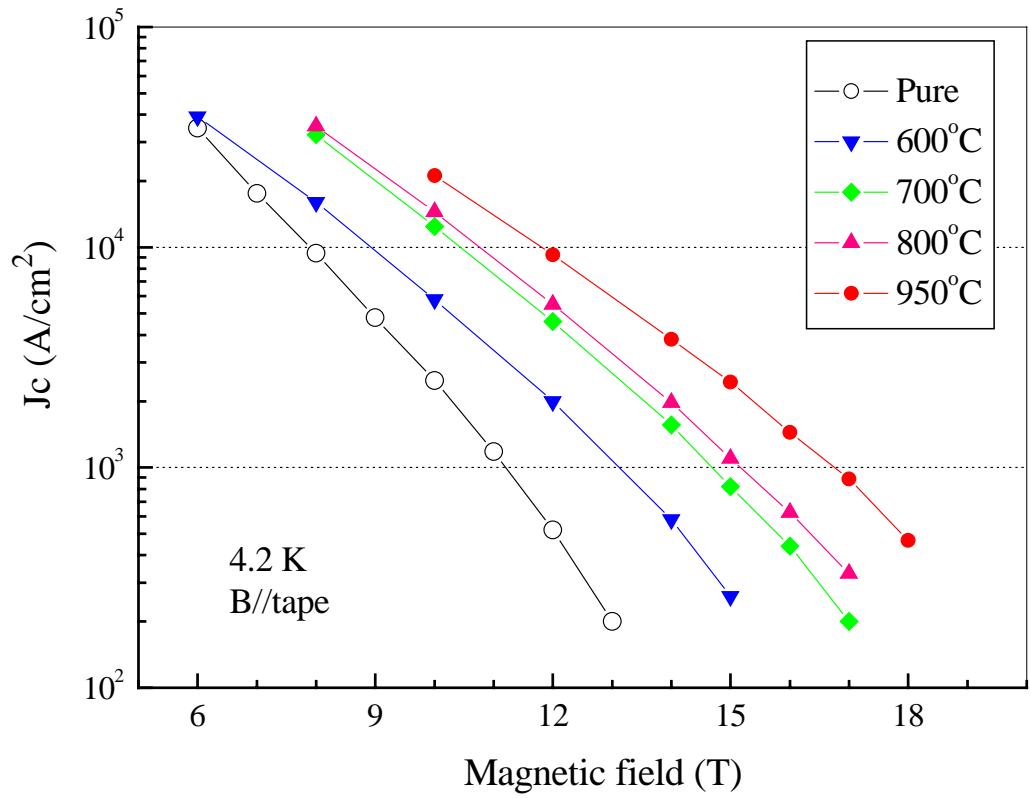

Fig. 1  Ma et al.

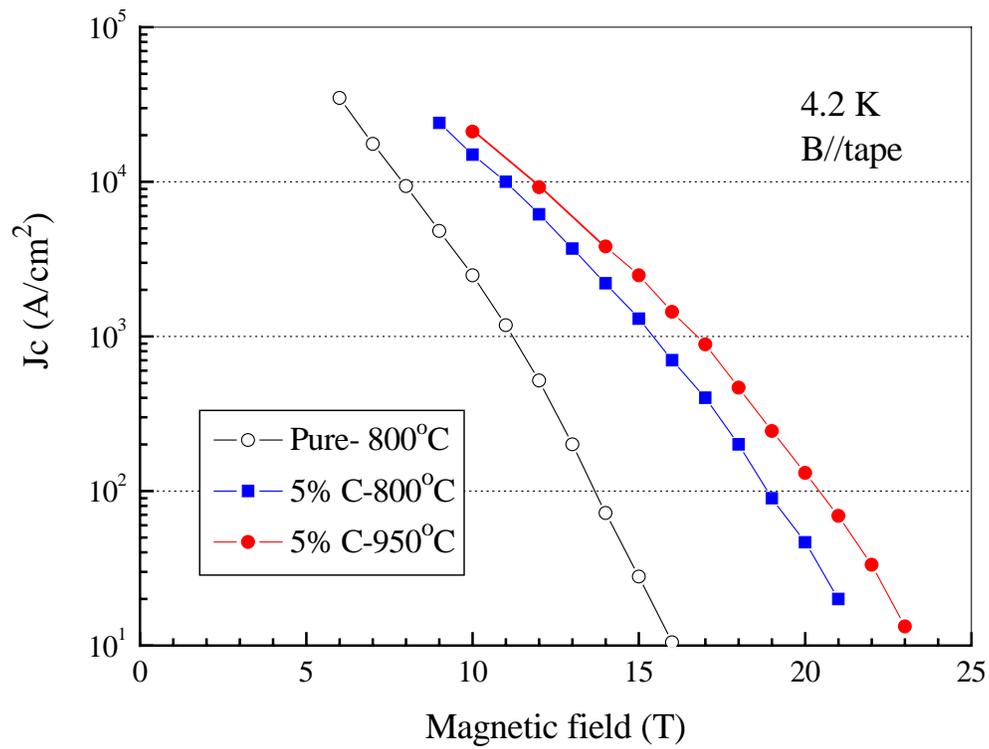

Fig. 2  Ma et al.

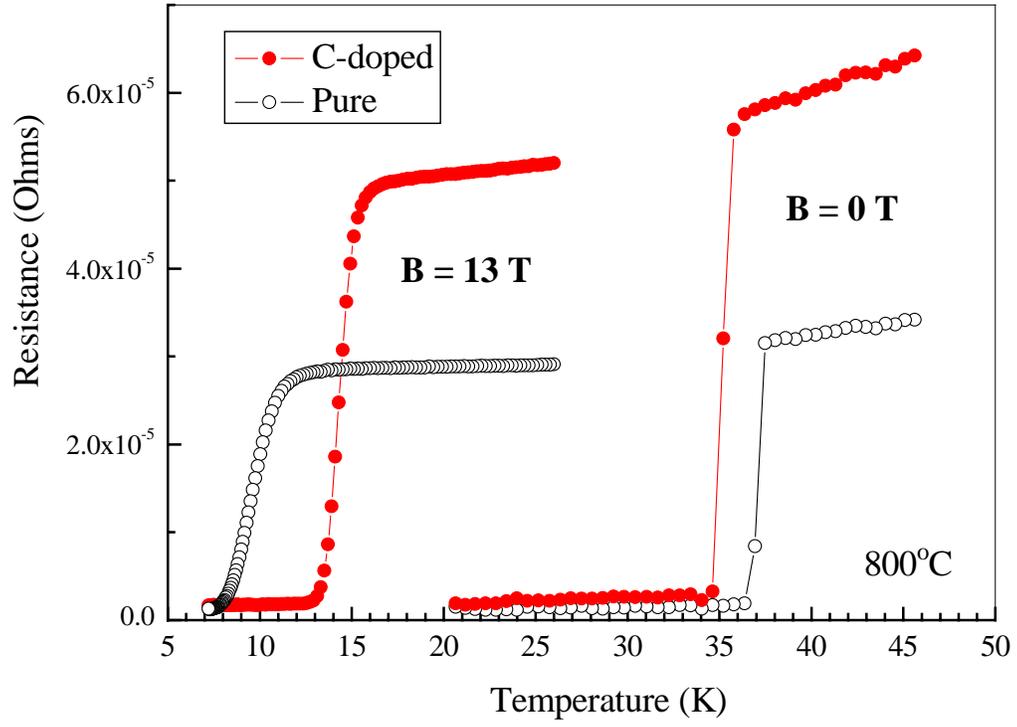

Fig. 3 Ma et al.

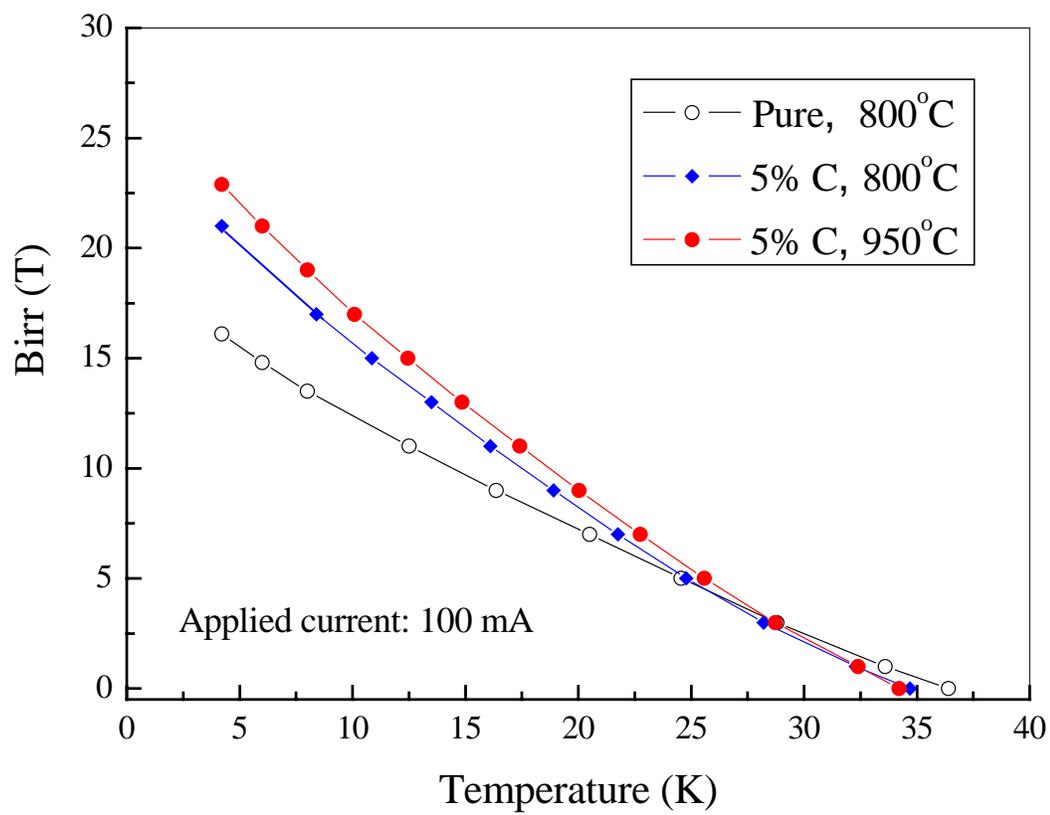

Fig. 4  Ma et al.